\documentclass[conference]{IEEEtran}
\IEEEoverridecommandlockouts
\usepackage{cite}
\usepackage{amsmath,amssymb,amsfonts}
\usepackage{algorithmic}
\usepackage{graphicx}
\usepackage{textcomp}
\usepackage{xcolor}
\usepackage{float}
\usepackage{wrapfig} 
\usepackage{makecell}
\usepackage{array}
\usepackage{xcolor}
\usepackage{url}
\usepackage{hyperref}
\usepackage{xurl}
\usepackage{enumitem}

\newcounter{researchq}

\def\BibTeX{{\rm B\kern-.05em{\sc i\kern-.025em b}\kern-.08em
    T\kern-.1667em\lower.7ex\hbox{E}\kern-.125emX}}
\begin{document}

\title{Federating Trust Perimeters: Extending Industry IAM with DLT-Based Governance}

\author{\IEEEauthorblockN{Carlo Segat}
\IEEEauthorblockA{\textit{Service-centric Networking} \\
\textit{Technische Universit\"at Berlin / T-Labs} \\
Berlin, Germany \\
carlo.segat@tu-berlin.de}}

\maketitle

\begin{abstract}

Digital systems are becoming more integrated, autonomous, and cooperative. AI agents, future mobile networks, and machine-to-machine economies point to one trend: spontaneous, cross-organizational, unplanned interactions between non-human entities (NHEs). Trust establishment for them remains an open problem. Federation is the natural candidate, but established approaches, from OpenID Federation 1.0 and SAML to Federated Identity Management, presuppose what this setting denies them: manual, ahead-of-time configuration and a common trust anchor, whether pre-established members or a shared provider. Trust domains must therefore federate without being prefigured to do so: plan for unplanned interactions. This paper examines whether prominent Identity and Access Management (IAM) approaches, namely SPIRE, Workload Identity Federation (WIF), and OpenID Federation 1.0, can support such federation. Drawing requirements from disparate fields (medical, mobile networks, agentic AI), it argues that SPIRE is the most promising starting point, but needs three extensions to meet them all: token exchange, letting a home domain mint scoped, audience-bound tokens from a foreign workload's SPIFFE Verifiable Identity Document (SVID); remote attestation, so a trust decision targets a specific workload rather than a whole domain; and a distributed-ledger layer that anchors trust roots, carries federation governance, and publishes the shared keys the other two depend on.

\end{abstract}

\section{Introduction and Motivation} 
\label{section:introduction}

The boundaries between digital systems are dissolving. Former silos, cleanly split into consumers and producers, are now interdependent in ways that challenge pre-established orders. Organizations collaborate not only through legal and business agreements but by integrating their systems technically, increasingly at the application layer (L7). One symptom is stakeholder proliferation within any single domain. This paper argues that tomorrow's digital systems must, by design, plan for unplanned interactions. Three trends in three separate domains illustrate the shift.

\textbf{5G and Beyond}
1G-era mobile network operators (MNOs) were vertically integrated, managing the system end-to-end \cite{cellularBook}. Successive generations decoupled infrastructure from services, pushing operators toward delegation and outsourcing, multiplying stakeholders and fragmenting the once-monolithic system. 5G was the milestone: cloud computing became the paradigm for network functions (NFs) and, through software-defined networking, for networking itself. The resulting heterogeneity is already visible: Telef\'onica runs non--real-time NFs (mostly analytics) in GCP \cite{googleTelefnicaCase}, and Amazon partners with Verizon and Vodafone to extend AWS to the MNOs' edge \cite{awsWavelength}.
Private networks tighten these relationships further. Their stringent requirements make the Service Level Agreement (SLA) contractually central: it encodes latency, reliability, and availability targets on which production safety can directly depend. Were those targets fixed at signing, the vertical would be a mere consumer of the MNO service. In practice they must be traded off against the dynamism of production processes and network conditions, a decision requiring domain knowledge the vertical may not want to share. Verticals therefore need a structural role in SLA governance, since the MNO cannot fully represent their interests.
Looking ahead, 6G is expected to build on device-to-device sidelink, letting user equipment relay traffic rather than only consume the network. As subscriber devices become, in part, infrastructure, the line between operator and user blurs \cite{ganesan2023sidelink}.
A unified access-control model for NFs within and across core networks remains active work, as a May 2026 3GPP S3 key issue shows \cite{3gpp-s3-262608}. Options borrowed from web practice exist, but open questions and configuration sprawl raise security concerns \cite{garzon2024beyondcertificates}.

\textbf{eHealth}
Digital health, or eHealth, is a broad term encompassing intersections of technology and medicine. Tracing the high-level history of digital health reveals the same trend of increased heterogeneity of stakeholders and collaboration.
A striking fact is that it was only recently that academia recognized the potential of patients themselves to directly improve the state of medicine. Examples include the e-patient white paper \cite{participatorymedicine} and the BMJ Patient Panel \cite{richards2013let}. Those are examples of how the idea of an unquestionable doctor taking unilateral decisions on behalf of a subordinated patient has been changing. Technology is the enabler for patients to emerge from their passive role. Conversely, technology is also the only plausible means of managing the new complexities that this collaborative relationship brings about.  
Although the medical domain has been traditionally slow and cautious in adopting innovation, recent trends show the dynamics may be changing. 
One example is the FDA Pre-Cert pilot (2017-2022), which explored a novel eHealth software lifecycle model intended to reduce per-release reviews \cite{fda2022precert}. 
Despite decades of progress, eHealth still lacks a single global data standard, with competing formats and interoperability constraints persisting \cite{ayaz2021fhir}. These problems compound across applications, such as practice management software, pharmacy and laboratory systems, and dental management tools, which are typically built by different vendors that do not always prioritize interoperability. The result is data silos, offering no complete view of a patient's clinical picture. Useful, but hard-to-integrate sources such as insurance claims and user-generated data from wearables and self-measurements compound the problem \cite{dinhle2019wearable}.
The eHealth ecosystem is judged as a fragmented one \cite{kffhealthnewsDeath1000} and among the many contributing factors to this state, one that is often cited is the lack of standardization in identity systems \cite{healthdatamanagementPatientMatching}, and the lack of incentive structure for data sharing \cite{rhapsodyReasonsInteroperability}.
The role of AI in the medical world is one of acceleration. Although diagnosis and drug discovery are often cited as the areas of strongest AI contribution \cite{thomas2025artificial}, others, including administrative efficiency and personalization are also relevant \cite{johnson2021precision, admin_e_health}. The 2018 introduction of an official AI policy by the American Medical Association marks a milestone in the integration of AI into the medical world \cite{crigger2019making}.

\textbf{Agentic AI}
The domain of artificial intelligence has shifted from being exclusively made up of specialized models privately managed to general-purpose and open ones adopted by a broader and more diverse public. 
An AI agent ecosystem is constituted by heterogeneous stakeholders by design. There are data owners, the source of the data that models are trained on~\cite{longpre2023dataprovenance}; infrastructure providers, whose role grows with the compute demands of modern models~\cite{sevilla2022compute}; chip manufacturers, exemplified by NVIDIA's rapid ascent~\cite{nvidia2025q4}; and finally users, whose interactions feed precious data back into it~\cite{arrietaibarra2018data}.
An AI agent's goals can be open-ended, so it cannot always anticipate the APIs and tools it will need, making collaboration with diverse parties essential \cite{ehtesham2025interopsurvey}.
To overcome scalability limitations such as those related to context windows or knowledge gaps, or to coordinate complex decisions, agents may also interact with each other \cite{google2025a2a}. In multi-agent systems (MAS), this can be confined to the boundaries of the same organization, but cross-organizational MAS are also foreseen. 


Agentic AI exacerbates the trust problems of the other domains: mobile and medical can bound their stakeholders to an extent, whereas an agent's participant scope is unbounded, under-specified goals and strategies being a distinctive feature.
The Agent2Agent (A2A) protocol targets cross-organizational use but defines no Identity and Access Management (IAM) of its own, advertising authentication schemes in signed agent cards (OAuth~2.0, mutual TLS, OpenID Connect (OIDC)) while leaving federation, authorization, orchestration, and context transfer to implementers \cite{csaA2AThreatModel}.
The Agent Network Protocol (ANP) goes further: Decentralized Identifier (DID) authentication via the \texttt{did:wba} method, a peer-to-peer model, dynamic protocol negotiation, and an Agent Description Document for discovery \cite{anp_whitepaper}. Still, it defines no formal authorization model; because \texttt{did:wba} and its \texttt{.well-known} discovery both resolve over HTTPS, they inherit the weaknesses of DNS and web public-key infrastructure (PKI) (e.g., DID-document spoofing, insecure hosting); and negotiating machine-level protocols in natural language introduces failure-prone non-determinism \cite{garzon2025aiagentsdecentralizedidentifiers}.
Open Finance is yet another example: regulation forces banks to expose customer data, opening the silo to a diverse crowd of third-party providers.

Across these domains, the same structural problems recur.

\begin{itemize}
    \item \textit{Disaggregation and heterogeneity.} Independent stakeholders and heterogeneous components rule out a common set of security assumptions; trust must be established explicitly and verifiably across technological boundaries.
    \item \textit{Security.} Unplanned interactions invite unforeseen failures and enlarge the attack surface.
    \item \textit{Ephemeral network position.} Cloud patterns make network locations unstable, undermining layer-3 and layer-4 identifiers as an IAM basis \cite{babakian2022retrospective}.
    \item \textit{Scale.} Not a few long-lived entities, but potentially millions of ephemeral ones.
\end{itemize}

Data silos have many causes, legal ones not least; a technical one is the absence of an automatic, transparent, and shared mechanism for establishing trust. Such frictions slow digital ecosystems moving toward broader and deeper application-layer integration. This paper contributes an analysis of established IAM solutions as the basis for designing an open, cross-organizational, flexible trust establishment mechanism for the digital ecosystems of tomorrow. None of them satisfactorily addresses trust-anchoring, or governance more generally, which is of particular relevance in a multi-stakeholder identity system: all rely on centralized or out-of-band coordination, leaving a single point of compromise and no tamper-evident record of federation membership. This gap, as this paper argues, can be appropriately filled by distributed ledger technology (DLT).

\section{Related Work}
\label{section:related_work}
\subsection*{SPIFFE \& SPIRE} \label{section:related_work:spire} SPIRE, the reference implementation of the SPIFFE specification, is a vendor-neutral, cloud-native framework for workload identity. It solves the problem of provisioning credentials to ephemeral and dynamically scheduled workloads in a secretless manner, enabling workload authentication. What sets SPIRE apart from other IAM solutions is the focus on operational automation. Once initial setup is performed, a dev-ops engineer can manage the entire lifecycle of thousands of certificates by means of a few configuration objects. SPIRE deployments scale by flexibly supporting many topologies (multiple Servers, hierarchies, etc.), including, importantly for this paper, federated ones \cite{spiffe_scaling_spire}.

\begin{figure}[t]
    \centering
    \includegraphics[width=1\linewidth]{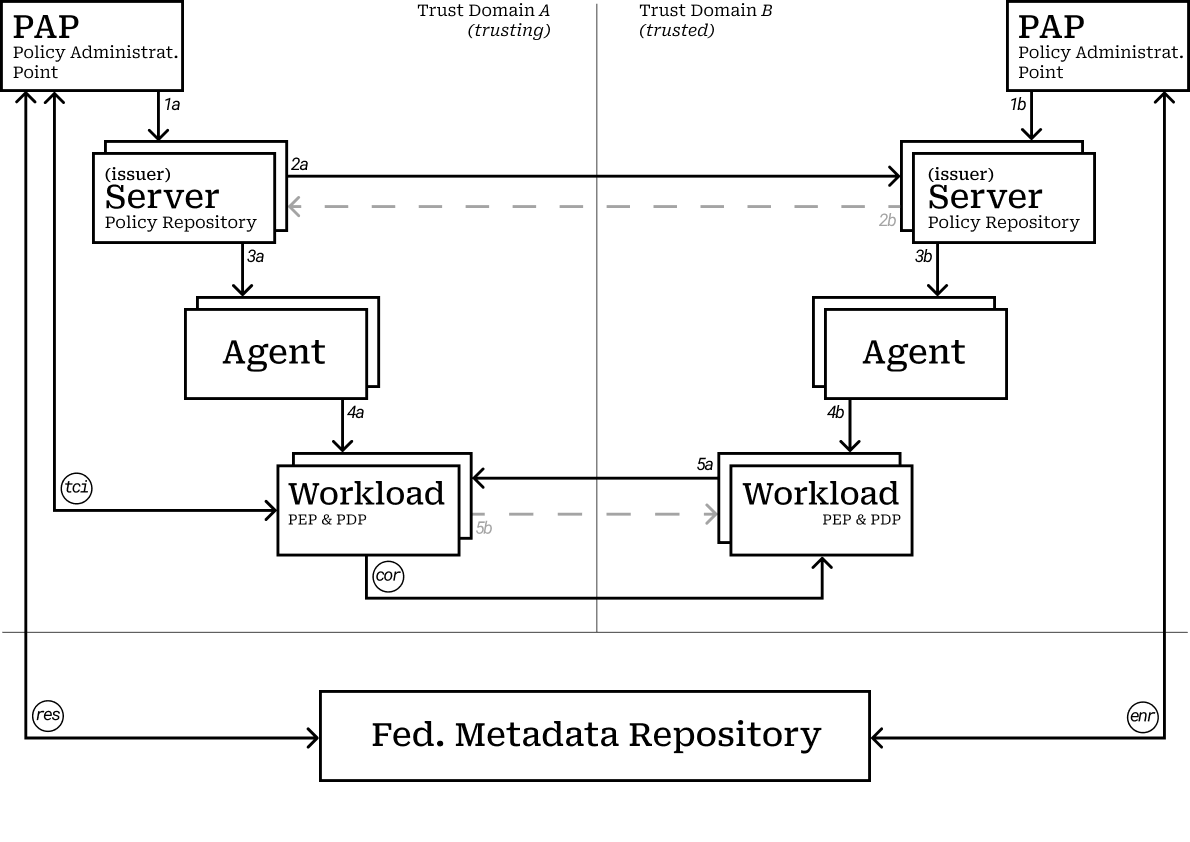}
    \caption{SPIFFE/SPIRE communication diagram showing a federated trust domain (right) and a consumer (left). Circled labels mark proposed enhancements. Dashed lines refer to interactions only required for the Imperative mode.}
    \label{fig:spire_arch}
\end{figure}

One half of Figure~\ref{fig:spire_arch} (with the exclusion of the circled interactions) illustrates the architecture of a SPIRE deployment within one trust domain. The Server acts as the issuer, signing the credentials (known as SVIDs) of all other entities. A Workload manages its SVID via the Workload API, which is exposed by the Agent. An Agent manages a number of workloads which must be co-located on the same host, because importantly, the Workload API is by design an unauthenticated interface. A Server establishes trust in its Agents via Agent Attestation, in which the Agent proves the host identity to the server (e.g., deployment-time secrets, cloud-provider-issued documents, or trusted platform modules). An Agent establishes trust in a Workload via Workload Attestation, in which the Workload's identity is proven. Concretely, the Agent interrogates the platform, querying the runtime (the operating system, Kubernetes, the container runtime) to check a workload's scheduling information.

SPIFFE strictly solves authentication. Authorization needs other techniques such as policy engines (e.g., Open Policy Agent) and separate permissions-carrying credentials.

\subsection*{OAuth} OAuth 2.0 is the de-facto framework for delegated authorization, defined by RFC~6749 \cite{rfc6749}. Over the years it has accumulated many extensions, such as token revocation (RFC~7009), token introspection (RFC~7662), dynamic client registration (RFC~7591), and token exchange (RFC~8693). Importantly, OAuth is not a tightly defined IAM approach on its own but rather a toolbox of endpoints, grant types, and token semantics from which concrete deployments are assembled. RFC~6749 itself admits it is ``likely to produce a wide range of non-interoperable implementations'' \cite[\S1.8]{rfc6749}. OAuth covers both human-facing and machine-to-machine (M2M) scenarios. Its general approach is the delegation of authorization: rather than sharing credentials, a resource owner (RO) authorizes a client to receive a scoped access token, which the client then presents to a resource server. Whereas the human flow issues a token only after the RO consents (by issuing an authorization grant), the M2M case uses the Client Credentials grant (RFC~6749, \S4.4): with no user to consent, no authorization grant is needed. The client is effectively its own resource owner and authenticates directly to the authorization server (AS) with its own credentials to obtain a token.

\subsection*{Workload Identity Federation} Workload Identity Federation (WIF) is best understood as an emerging, cloud-vendor-driven practice rather than a single standard. The term and the workload-centric practice were introduced by Google Cloud in 2021 \cite{gcp_wif} and have been adopted by other providers. WIF's goal is to let a workload, often running outside the target cloud (on-premises, in another cloud, or in a CI/CD pipeline), access cloud resources without long-lived service account keys \cite{gcp_wif}. 
WIF reuses building blocks from both OAuth and OIDC. Token exchange (RFC~8693 \cite{rfc8693}) is the central piece: the workload presents a credential from its own issuer to a Security Token Service (STS), which validates it and returns a security token that can be used to directly access a protected resource. Figure~\ref{fig:wif_arch} illustrates how Workload B asks the STS in A for a token (step 3); the STS fetches B's public keys to validate the request (step 4); once the STS has minted and returned a token, Workload B uses it to initiate communication with Workload A (step 5); Workload A may, if not cached, fetch the keys of the local issuer to validate the token sent by B (step 6). Although for simplicity the paper assumes OIDC, the type of credential is flexible. Google Cloud, for example, supports SAML signed assertions and X.509 client certificates, among others \cite{gcp_wif}. WIF's trust model is bilateral: each external issuer must be registered individually with the STS for its tokens to be accepted, a manual, per-issuer step.

\subsection*{OIDC and OpenID Federation 1.0}
OIDC Core 1.0 \cite{oidc-core} is an identity layer on top of OAuth 2.0 adding authentication. Its core construct is the ID Token, a JWT in which an OpenID Provider (OP) asserts a subject's identity to a Relying Party (RP) through a set of mandatory claims. OIDC also defines a discovery mechanism: the OP publishes its provider metadata as a JSON document at a well-known endpoint, advertising among other entries the URL of its JWKS Endpoint, from which the RP fetches the public keys that verify ID Token signatures. 
While OIDC's authentication ceremonies are designed around human logins and browsers, its model can be applied to M2M, where the subject is, for example, a workload and the OP its issuer. Nonetheless, OIDC's trust model is bilateral: each RP-OP pair must be set up manually, which does not scale across domains. OpenID Federation 1.0 \cite{openid_federation} improves this by making the model multilateral and, to an extent, by automating trust establishment via OIDC's very own hierarchical PKI. The communication diagram in Section~\ref{section:appendix_oidc} visualizes the flow. In the OIDC PKI, all entities publish signed Entity Statements: self-issued JSON documents containing metadata about the entity. Intermediate authorities (known as Intermediates) additionally publish Subordinate Statements (i.e., a list of leaf entities they vouch for). Trust Anchors (TAs) sign the Intermediates' Entity Statements and must be pre-trusted by clients. The fact that trust can be chained back to a commonly trusted TA allows OPs to trust unregistered RPs, a mechanism formally known as Automatic Registration. 
Trust Marks may further attest an entity's conformance to a federation profile. 
As stated, the goal of OIDC is for an RP to obtain a subject's identity information from the subject's OP. This is the case for some prominent use cases of OpenID Federation 1.0. For example, in the EU Digital Identity Wallet ecosystem (in production in Italy), an RP obtains a citizen's attributes from an identity (wallet) provider anchored to a national TA \cite{spidcie_oidf}. In open banking, a bank (OP) releases a customer's account data to a third-party provider (RP) only after resolving that provider's chain to the scheme's TA (a model championed by Open Finance Brasil \cite{raidiam_oidf}). 

\subsection*{Organizational Culture} Two different groups stand behind the standards associated with OAuth and OIDC: the OAuth Working Group, stewarded by the IETF, and the OpenID Foundation. Substantial overlap exists not only between the problems they address, but also between their members (e.g., Nat Sakimura is an author of both OpenID Connect Core 1.0 and various OAuth-related RFCs). Nonetheless, differences exist. The OpenID Foundation tends to be more identity- and authentication-centric, focusing on login flows and sessions. It also has a more commercial attitude, financing itself through membership fees, certification programs, and member-funded projects \cite{oidf_about}. In OAuth the focus is more on authorization and APIs, with the fundamental problem of how to delegate authorization at the center. Being part of the IETF, membership is free and financing relies on donations. SPIFFE and SPIRE do not come from a standards body but from an open-source project in which the specification (SPIFFE) and the reference implementation (SPIRE) evolved together. It is a community-driven Cloud Native Computing Foundation project whose cultural center of gravity is cloud-native infrastructure and zero-trust workload identity.

\subsection*{Token Exchange} Peer-to-peer validation, not mediated by an exchange process, is natural in OIDC. This is because since its inception, OIDC standardized the token format (JWT) and metadata/key discovery (OIDC Discovery). Those two aspects were not tamed by OAuth, which for nine years left implementers free to choose their token format and was perfectly happy with opaque tokens that could only be AS-validated; the JWT Profile for OAuth Access Tokens (RFC~9068) appeared only in 2021. Moreover, only in 2018 did OAuth adopt a metadata discovery mechanism (RFC~8414) very similar to the OIDC one. Because OIDC mandated both token format (JWT) and mandatory claims (e.g., \texttt{iss}, \texttt{sub}, \texttt{aud}) the need to exchange tokens was not a priority; RPs and OPs could understand each other's tokens and could validate them directly. On the contrary, exchanging tokens became a necessity for OAuth, where token-interoperability between domains was not provided by design. RFC~8693 standardized the STS concept, allowing a home domain to issue tokens in its own format to entities in a foreign domain. The interoperability problem still existed, but it was confined to the STS rather than remaining a system-wide concern.
Token exchange has implications for the trust and operational model. Exchanging tokens gives the domain owner full control over its trust relations. Another advantage is that adding a new trusted issuer does not require downstream changes. The cost is that of managing the STS and potentially many heterogeneous policies. As WIF makes the case for it, token exchange is an IAM feature that facilitates bilateral trust relationships. Many-to-many trust models, such as those between members of SAML-like federations (e.g., eduGAIN), do not need token exchange. In those cases, there is a greater technical minimum common denominator agreed upon beforehand by federation members. In other words, all members ``speak the same IAM language'' and a translation layer is not needed. This can also be observed in OpenID Federation 1.0 where tokens are used directly. 

\subsection*{Remote Attestation}
SPIRE's attestation is local by construction: the agent and the workload it attests are assumed to be colocated~\cite{spiffe_concepts}. Attesting a workload across a trust-domain boundary is instead the subject of remote attestation, whose terminology the IETF's Remote ATtestation procedureS (RATS) architecture standardizes in RFC~9334~\cite{rfc9334}: an Attester produces Evidence about itself, a Verifier appraises that Evidence against references and produces Attestation Results, and a Relying Party consumes those results to decide. Implementations exist for both roles: Keylime, a CNCF project, uses TPM~2.0 to attest boot and runtime integrity of remote nodes~\cite{keylime}, and Veraison, hosted by the Confidential Computing Consortium, provides a scheme-agnostic Verifier~\cite{veraison}. Work to carry attestation across identity boundaries is emerging: an experimental plugin makes a node's Keylime attestation state a precondition for SPIRE issuing identities on that node~\cite{keylime_spire}, and two unadopted Internet-Drafts targeting the IETF's Workload Identity in Multi System Environments (WIMSE) Working Group propose carrying RATS Evidence inside a Workload Identity Token~\cite{wimse_wit_attestation} and chaining hardware evidence transitively to an SVID~\cite{wimse_transitive}. In all of them the relying party's trust in the appraising authority, be it a Verifier or the endorsement chain behind a hardware root, is bilateral and pre-configured: RFC~9334 reduces it to holding the Verifier's key in a trust anchor store, or to having the Verifier first act as an Attester towards the relying party~\cite[\S7.1]{rfc9334}. What is missing is a multilateral mechanism by which mutually distrusting domains agree on which Verifiers are acceptable; a recent individual draft sketches decentralized trust registries for this purpose but stops at architecture, specifying neither protocol nor governance~\cite{rats_federated_infra}. That is the gap the ledger layer of Section~\ref{section:iam_and_designs} addresses.

\subsection*{SAML Federation}
The Security Assertion Markup Language (SAML) 2.0 is the long-standing standard for cross-organizational web single sign-on: an identity provider (IdP) signs an assertion vouching for a subject and a service provider (SP) consumes it, so one login reaches services across organizational boundaries. Trust rests on a pre-established circle of IdPs and SPs exchanging signed metadata ahead of time, which in large inter-federations (e.g., eduGAIN) a central operator aggregates and distributes. SAML thus exemplifies the closed-world federation this paper sets out to relax.




\section{Requirements and Research Questions}
\label{section:requirements}

The following requirements describe a future cross-domain trust establishment framework that fills the gaps and realizes the potential discussed in Section~\ref{section:introduction}. 

\begin{itemize}
  \item \textbf{R1}: unplanned interactions. A workload in domain X must talk to a workload in domain Y without prior pairwise configuration;
  \item \textbf{R2}: non-human interaction. No assumptions of interactive consent or browser redirects;
  \item \textbf{R3}: ephemerality. Workloads come and go, and identity must be decoupled from network locators;
  \item \textbf{R4}: heterogeneity. Trust domains may use different IAM solutions;
  \item \textbf{R5}: scale. Federation size grows to millions of non-human entities (NHEs);
  \item \textbf{R6}: fast trust-change propagation. Federation on-boarding,
  off-boarding, and key rotation must converge across federation members and
  consumers within bounded time;
  \item \textbf{R7}: least-privilege authorization. Access is scoped; trust does not imply blanket access;
  \item \textbf{R8}: auditable governance. Federation membership and services must be inspectable by all members and consumers;
  \item \textbf{R9}: no single point of failure. No single party can unilaterally censor or partition the federation.
\end{itemize}


The first research question follows.

\phantomsection\label{rq:one}
\textbf{RQ1}. Which IAM approach is the best foundation to extend toward dynamic federation establishment between trust domains, satisfying the requirements outlined in
Section~\ref{section:requirements} and enabling access control (i.e., authentication and authorization) between workloads of different domains?

\hyperref[rq:one]{RQ1} is answered analytically. The requirements eliminate SAML and OpenID Federation 1.0, leaving SPIRE and WIF. What separates the two are two aspects that are hard to benchmark. The first is trust granularity. SPIRE roots trust in attestation of the specific workload, whereas WIF can only verify the assertions carried inside an exchanged token. The second is standardization coverage. SPIFFE/SPIRE specifies its own internals (the Server, the Agent, and how workloads are attested and provisioned), whereas WIF leaves provisioning to ambient, infrastructure-specific mechanisms and exchange-policy encoding to each provider (Section~\ref{section:iam_and_designs}). The empirical evaluation therefore does not arbitrate \hyperref[rq:one]{RQ1}. It validates the resulting design and measures WIF alongside it as an industry baseline.

Because none of the considered IAM solutions satisfactorily supports the trust-anchoring and governance requirements (R8, R9), a second, separate research question follows.

\phantomsection\label{rq:two}
\textbf{RQ2}. How can a multi-stakeholder coordination mechanism between heterogeneous trust domains be designed to provide trust anchoring and metadata sharing?

\section{IAM Solutions and Proposed Design} 
\label{section:iam_and_designs}
Among the IAM solutions surveyed in Section~\ref{section:related_work}, SAML federations are not considered further. They are browser-centric, which R2 (non-human interaction) rules out, and presuppose a pre-established circle of members, which R1 (unplanned interactions) rules out. They also carry heavy member-management overhead, rest on older XML conventions, and in general are not designed for M2M APIs.
On similar grounds, OpenID Federation 1.0 is also excluded. Any two domains wishing to interact must already have joined the federation, a manual on-boarding step, and must share a TA. This breaks R1 and R2 as well. The distinguishing feature of OpenID Federation 1.0 is precisely the PKI it defines, which this paper proposes to relocate onto a ledger. By contrast, SPIRE and WIF are less opinionated about how trust is rooted: each contributes an orthogonal primitive (attestation and token exchange) that leaves more control over how the anchors are realized. Both also bind identity to the workload rather than to its network location, as R3 requires.

The diagrams below show how SPIRE and WIF realize the discussed dynamic and unplanned trust establishment.
Arrowheads point from initiator to responder (data may flow both ways); circled labels mark the non-standard interactions this paper proposes; components carry their federation-level authorization role per NIST SP 800-162~\cite{nist-abac}, local roles being omitted.
A domain can have three, possibly concurrent, roles in a federation: member, provider, and consumer, with the sole constraint that a provider must be a member. In the diagrams, Domain B is a provider and Domain A is a consumer. 
There are two operational modes in which A can consume trust: Imperative and Inversion of Control (IoC).
Under Imperative, the consumer invokes services in the federated/trusted Domain B. Under IoC, it is the federated/trusted domain that invokes services in the consumer. The diagrams visualize both modes simultaneously, disambiguating with dashed lines for the Imperative mode. 
The Policy Administration Point (PAP) is defined by NIST as a policy management interface \cite{nist-abac}. As such, a principal (human or NHE) configures its relationship with respect to a federation (e.g., being a member or a consumer) by interacting with it.
The publicly available federation metadata is where a provider declares the terms under which federated workloads may access its resources. For the IoC case, it is the sole configuration source for the consumer. The metadata must also specify whether imperative mode is supported and how a consumer gains authorization (e.g., federation members only, or any holder of a credential from a given issuer), the required request parameters (e.g., the \textit{aud} and principal of an exchange request), and any external coordination procedure (e.g., payment settlement or interactive know-your-customer checks).
Wherever the PAP configures a workload, the configuration may equivalently target an indirection layer: environment variables, or a local discovery service queried at runtime.

\subsection*{Interactions Common to Both}
\begin{itemize}
    \item[enr)] Domain B enrolls in the federation; its metadata enters the Repository.
    \item[res)] Via the PAP, the federation metadata is resolved and configures Domain A: its workloads learn which foreign domains exist and what they offer, and B's workloads become authenticable and authorizable under IoC.
    \item[tci)] Trust Configuration Interface. Via the PAP, the foreign-domain URLs known to A's workloads are managed, enabling discovery as well as controls such as blacklisting. In the most extreme, agentic realization, this arrow instead has Workload A configure its own domain's trust perimeter bottom-up, establishing technical trust in a previously unknown federation and breaking with the closed-world model in which joining a federation is a static decision.
    \item[cor)] Coordination between workloads: how Workload A requests that Workload B operate under IoC, and how Imperative-mode authorization is negotiated.
\end{itemize}

\subsection*{SPIFFE \& SPIRE}
Although the trust relationship itself is asymmetric, whether the bootstrap must be symmetric, i.e., whether the two domains must pre-exchange bundles out of band, depends on the bundle endpoint's \textit{endpoint profile}. The deciding factor is whether TLS to the bundle endpoint relies on web PKI. With \textit{https\_web}, the endpoint presents an ordinary X.509 certificate rooted in a public certificate authority (CA) (e.g., Let's Encrypt); a consumer can therefore validate the connection and fetch the bundle through web PKI alone, with no prior knowledge of the provider. SPIRE is then fully asymmetric: the provider needs no configuration about its consumers. With \textit{https\_spiffe}, the endpoint instead presents its own X.509-SVID, which is not anchored in a public CA. Validating it, therefore, requires already holding the bundle, a chicken-and-egg problem: a domain must obtain the foreign bundle out of band before it can fetch it. Bootstrap is thus symmetric, and both domains must be mutually configured.

\begin{figure}[t]
    \centering
    \includegraphics[width=1\linewidth]{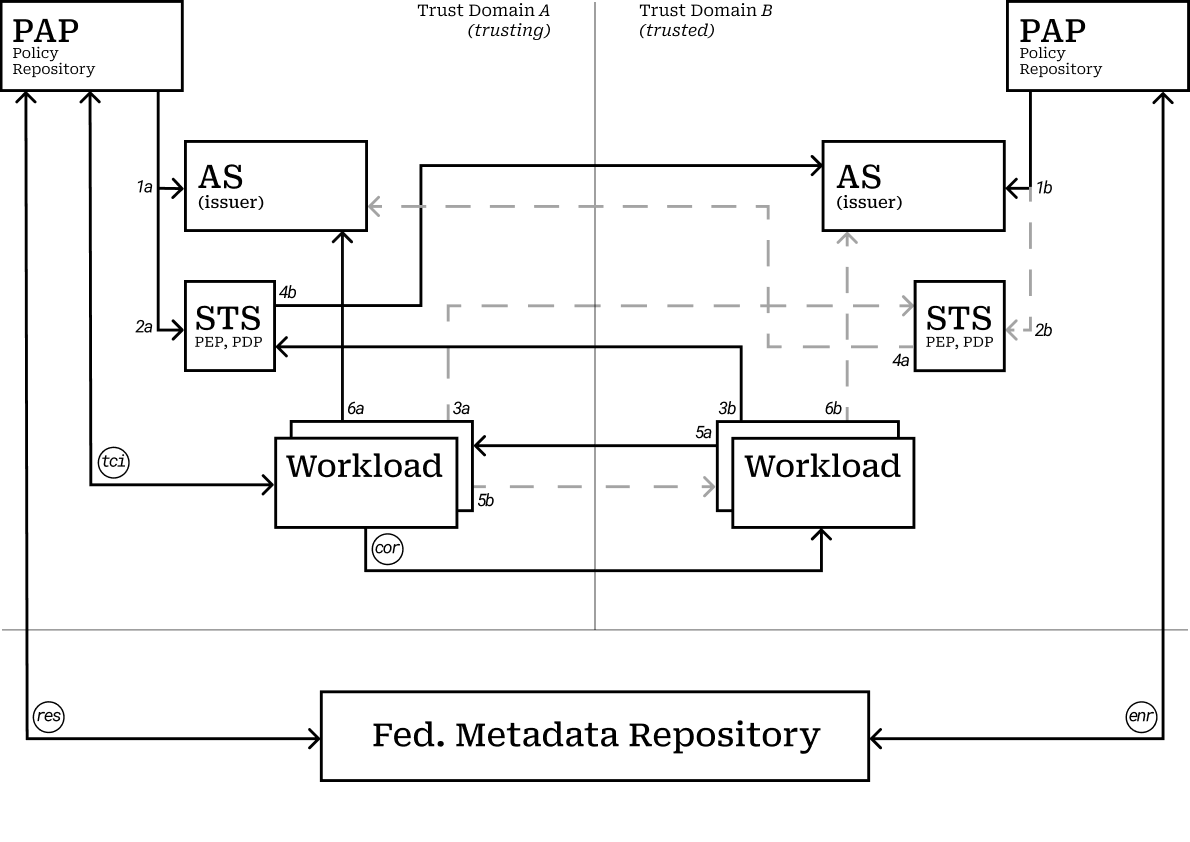}
    \caption{WIF communication diagram showing a federated trust domain (right) and a consumer (left). Circled labels mark proposed enhancements. Dashed lines refer to interactions only required for the Imperative mode.}
    \label{fig:wif_arch}
\end{figure}

\begin{enumerate}
    \item[1)] The Server is bootstrapped (configured parameters include local domain name, node-attestations, workload registration entries, etc.). Under \textit{https\_spiffe}, configured parameters must also include the other server's bundle. 
    \item[2)] The Server fetches the bundle of the other domain. This must happen under \textit{https\_web}; under \textit{https\_spiffe} it is done to refresh a potentially stale bundle. 
    \item[3)] SPIRE Agent Attestation as per Section~\ref{section:related_work:spire}.
    \item[4)] SPIRE Workload Attestation as per Section~\ref{section:related_work:spire}. In SPIRE, the Agent pushes local trust bundle updates to the workloads. This is in contrast to WIF where workloads have to fetch from the local JWKS endpoint (cf. arrow 4 of Figure~\ref{fig:wif_arch}).
    \item[5)] Workload communication. Workload A acts as both enforcement and decision point (configured with the authorization rules from the Agent). Workload B initiates under IoC; Workload A under Imperative mode.
\end{enumerate}

\subsection*{WIF}

Whereas SPIRE's \textit{https\_spiffe} profile permits opting out of web PKI, WIF makes this dependency inescapable. The advantage is that mutual configuration between consumer and provider is never required. Note that the separation between STS and AS (i.e., the issuer) is logical, and deployments often combine them into a single component.

\begin{enumerate}[label=\arabic*)]
    \item[1)] The AS is bootstrapped with essential data such as keys and the domain name.
    \item[2)] The STS is configured with trusted issuer URLs, their configurations, and per-issuer policies for token exchange. In this design, the STS configuration is derived from the federation metadata. Nonetheless, domain A remains sovereign and may modify its STS policies as it sees fit. How policies are encoded remains an open design question, as RFC~8693, while standardizing the token-exchange protocol itself, does not prescribe how to encode exchange authorization policies (e.g., AWS uses a JSON IAM trust policy, GCP uses attribute conditions written in the Common Expression Language \cite{aws-trust-policy,gcp-wif-pool-provider}). 
    \item[3)] Workload B makes a token exchange request to A's STS. The request is evaluated against the configured policy. If successful, a token signed by Domain A AS is released.
    \item[4)] After receiving an exchange request, the STS checks whether it originates from a trusted domain by inspecting the issuer URI of the presented token. Verifying the token's signature, however, requires the issuer's public key, which the STS may need to fetch from the requesting domain's JWKS endpoint.
    \item[5)] After successfully exchanging a token, the workload can communicate with the other domain's workload, which, effectively, in the role of a Resource Server (as defined by RFC~6749), validates the received token against the local AS keys.
    \item[6)] A Workload fetches from the local JWKS URI the AS public keys. The JWKS URI is a \textit{well-known} endpoint derived from the issuer URL. The diagram omits how this URL is learned. In most cases, it is derived from the \textit{iss} field of the token the workload uses to authenticate with the AS (which, as stated before, is in most deployments obtained through ambient provisioning).    
\end{enumerate}

A noticeable difference from SPIRE is the amount of cross-domain interactions required. In SPIRE it was sufficient for the servers to fetch their respective bundles, but in WIF, the dance is more convoluted. The workload first calls the foreign STS (token exchange), then the foreign STS may fetch the key material from the requesting Issuer (if not cached). 

\paragraph{Workload Ambient Provisioning}
The diagram (and every OAuth and OIDC specification for that matter) assumes that the workloads are already provisioned with credentials, recognized by the local issuer, attesting their identity. Most deployments rely on the infrastructure to act as the Identity Provider (i.e., \textit{infrastructure as IdP}); Kubernetes projected tokens~\cite{k8s-sa-token-projection} and link-local metadata endpoint (e.g., \url{http://metadata.google.internal/})~\cite{gcp-metadata} are common examples. In some documents, this token is referred to as \textit{ambient token} because it is provisioned by the environment (i.e., the infrastructure) on which the workload runs~\cite{anthropic-wif}. For this reason, the paper refers to this practice as \textit{ambient provisioning}.

\subsection*{Federation in SPIRE and WIF} The defining difference in how SPIRE and WIF federate is token exchange. In WIF, a workload uses its local credential to obtain a target domain credential from the STS. Federation is thus mediated by a broker that decides how to map a foreign principal onto a local one. This token-exchange model forces the configuration in WIF to be two-sided and asymmetric. On the trusting side, the PAP configures a \textit{trust policy} on the STS declaring which foreign issuers and claims to accept and how to map them onto a target principal. On the trusted side, the PAP configures a \textit{federation client} that mainly carries the audience (the \textit{aud} that will be requested) 
SPIRE federation, by contrast, involves neither a token exchange nor an STS. Trust domains exchange only \textit{trust bundles}, the public keys (X.509 CA certificates and JWKS) required to verify a peer's SVID~\cite{gilman_harding_2021_spiffe}. There is no identity translation: a workload keeps its own SPIFFE ID across the boundary, and the consumer verifies the peer directly over mTLS against the federated bundle. Configuration is therefore symmetric: both servers publish a bundle endpoint and configure the other domain's endpoint.
SPIRE federation trusts an entire domain. WIF, by contrast, conditions its trust policy on issuer and claims, enabling fine-grained authorization. This expressiveness comes at a cost: the relationship is asymmetric and requires coordination between both parties.

\subsection*{Proposed Design: Attested Cross-Domain Identity} Composing SPIRE with WIF achieves access granularity, but at a cost: it abandons attestation exactly where it matters most. Neither gives a fine-grained decision anchored in attestation of the specific remote workload.
The limit is structural. Attestation requires observing the workload, which only a co-located agent can do; a token exchange offers no such observation, so the consuming domain merely verifies the assertions contained in the token. In contrast, the trust in an SVID is rooted in attestation.
Composition also adds overhead: two trust planes, different credential formats and their mapping to operate, plus WIF's asymmetric, provider-specific configuration whose two sides must be coordinated.
This paper proposes to extend SPIRE directly. A standardized token-exchange step, reusing RFC~8693~\cite{rfc8693} bound to SPIFFE semantics, would bring WIF's expressiveness (conditioning on claims, mapping to a local principal, scoping to an audience) into the model, bridging domains that run different IAM solutions (R4) and keeping cross-domain access least-privilege (R7). Remote workload attestation would let a trust decision target a specific workload rather than a whole domain. Together, they yield cross-domain identity that is both fine-grained and attestation-based, under one credential and one open protocol.
Both the token-exchange and remote attestation additions presuppose shared trust anchors. In a federation of potentially mutually distrusting domains lacking a common authority, this shared reference cannot be owned by any single member without making it a trusted third party with unilateral decision power, which is what R9 forbids.
For this reason, shared key material and federation metadata are recorded on the ledger. This solves key exchange and makes updates tamper-evident and consistently readable by all members and consumers, as R8 requires. The federation lifecycle operations, such as on-boarding new members, become a multi-stakeholder access-control problem, which is solved by a dedicated governance layer, such as that described in  \cite{2025programmable}. Verifier trust anchors are part of that shared key material, and admitting or removing a Verifier is a governance action like admitting a member. A domain can therefore appraise Evidence produced in a foreign domain against a Verifier the federation collectively accepted, rather than one it configured bilaterally and out of band. These records change infrequently, so the ledger stays on the control plane, off the runtime path of exchange and attestation.

\section{Evaluation Design}
This evaluation does not select the IAM approach. \hyperref[rq:one]{RQ1} is settled analytically in Section~\ref{section:iam_and_designs}: automatic, bottom-up workload attestation and a standard that also covers provisioning single out SPIRE as the base to extend. The evaluation instead validates that the extended SPIRE design meets R5 and R6, with WIF measured alongside as an industry baseline. At the same time, it answers \hyperref[rq:two]{RQ2} by evaluating the Federation Metadata Repository (FMR), which carries trust anchoring and governance, in its two forms: a centralized web server versus a distributed ledger.

The plan takes two perspectives, that of federation members and that of external consumers, and applies three measurement categories to both. The FMR is assumed to notify members of lifecycle events via Server-Sent Events when implemented as a web server, or via Solidity events when implemented on a ledger.

\textbf{Federation Operations.} The latency of lifecycle operations (federation creation, member addition, key rotation, and member removal) is measured against federation size, to rank the candidates and expose their scaling properties (under SPIRE, for instance, each foreign domain adds an SVID field, anticipating increased overhead as federation size grows). Key rotation
adds measurements of the internal propagation time (how long it takes for the local workload to pick up the rotated key), external propagation (how long it takes for foreign domains to do so), and full-mesh latency (how long it takes for the workloads affected by the key rotation to communicate with every other foreign workload using the new key).
Full-mesh latency is asymmetric: under SPIRE, it is the time for the receiver to re-establish its federation configuration after a sender-side rotation, whereas under WIF, it is the time for the sender to re-exchange tokens after a receiver-side rotation. 
Member removal splits into propagation latency and empty-mesh latency (how long it takes for the removed member's workloads to be unable to reach anyone).

\textbf{Workload Under Load.} Per-request validation should be constant, but shared resources on the validation path (introspection endpoints, JWKS caches, locks) may degrade non-linearly under load. With $N$ senders driving one target, sender-side latency is plotted against load. A curve plotted with validation disabled fixes the setup's baseline, and a curve with validation enabled sits above it, the gap being the validation cost. The measurement establishes whether that gap stays constant before it widens, and which candidate is cheaper. Separate setups are constructed to measure WIF with and without token introspection.

\textbf{Active/Passive Revocation Trade-off.} With token exchange, the exchanged token's TTL serves as passive revocation; active revocation would instead introspect every request at the STS, at a cost. Throughput (application messages per second) is plotted against TTL. The active curve introspects on every request and is TTL-independent, a horizontal reference; the passive curve does not introspect and sweeps the TTL down, starting above the reference (as it does not bear the introspection overhead) but falling as shorter TTLs raise the re-exchange rate. Their crossing is the optimal TTL: the tightest TTL matching active-mode throughput. This applies only to WIF today; it would extend to SPIRE if SPIRE adopted token exchange.

\section{Conclusion and Future Work}
\label{section:conclusion}
This paper claims that the future digital landscape will see deeper technical integration between systems. Because of this, and additionally motivated by the rise of large-language-model-based AI agents, supporting unplanned interactions between NHEs will become more relevant. 
The approach explored by this paper is to evolve the federation-building capabilities of today's IAM solutions to enable a single trust domain to federate under loosely coupled groupings that agree and coordinate on the essential trust parameters. 
After analyzing the most prominent IAM solutions, this paper answers \hyperref[rq:one]{RQ1}: SPIRE is the most promising foundation to extend. Nonetheless, to satisfy the need for dynamism, automation, and control in trust-relationship management, SPIRE should be extended with remote attestation, token exchange, and, in answer to \hyperref[rq:two]{RQ2}, a distributed ledger acting as the Federation Metadata Repository, anchoring the shared trust roots and carrying the governance actions that admit and remove members without any single domain owning that reference.

This paper presents a design, not an implementation. Building it end-to-end remains to be done, and the work is substantial. Four issues stand out: SPIRE remote attestation, SPIRE token exchange, the distributed-ledger layer that anchors trust roots and carries governance, and, not least, revocation.
Revocation is the most tractable of these. SPIRE relies on short SVID lifetimes, letting credentials lapse rather than recalling them. Routing issuance through the STS adds finer control: a token can be revoked directly instead of waiting out its TTL.
Token exchange must be brought into SPIRE, introducing an architectural element functionally equivalent to WIF's STS to mint scoped, audience-bound tokens from a foreign workload's SVID.
The ledger layer likewise remains to be realized. Trust-root anchoring and the governance actions that admit or remove domains and Verifiers were described as ledger-backed, but the concrete choices are open: which ledger, which consensus algorithm, and which access-control model.
Remote attestation is perhaps the most arduous task. SPIRE assumes Agents and Workloads are co-located. Remote attestation does not escape this. Transitive trust cannot be eliminated but only relocated; the guiding principle is to shift the basis of trust from an issuer's assertion to attestation evidence produced by the workload itself.
Practical directions include using a hardware anchor such as a trusted platform module, or delegating appraisal to a Verifier in the RATS sense. Which Verifier a domain accepts is not left to bilateral configuration: its trust anchor is recorded on the ledger and admitted through the same governance layer that admits members, so what remains open is the attestation machinery itself rather than the question of whom to trust to appraise it. 

\section*{Acknowledgments}
This work was funded by the Federal Ministry of Education and Research (BMBF) in Germany under the grant number 16KIS2251 of the SUSTAINET\_guarDian project.


\clearpage

\appendix
\label{section:appendix}

\subsection*{OpenID Federation 1.0}
\label{section:appendix_oidc}

\begin{figure}[t]
    \centering
    \includegraphics[width=1\linewidth]{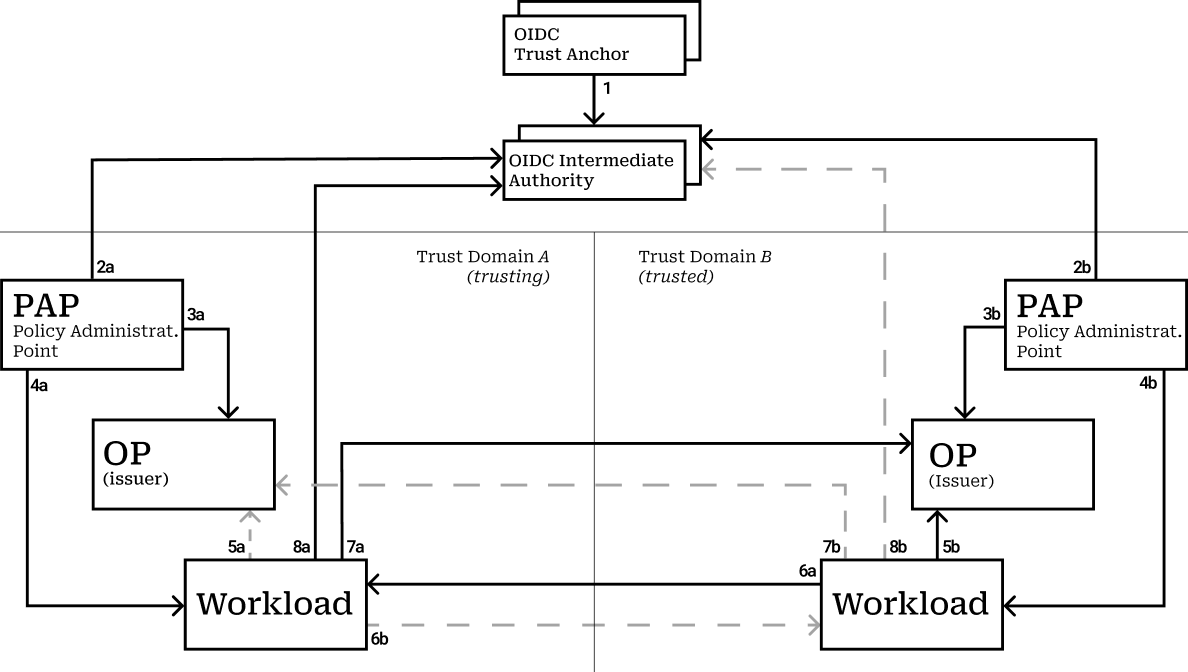}
    \caption{OpenID Federation 1.0 communication diagram showing two federated trust domains where the right one initiates communication with the left one. Dashed lines refer to an interaction initiated by the left domain. OpenID Federation 1.0 requires bilateral membership (i.e., both A and B are members of the federation).}
\end{figure}

\begin{enumerate}[label=\arabic*)]
    \item[1)] A TA publishes its own self-signed Entity Configuration and Subordinate Statements defining its immediate subordinates (i.e., the intermediates).
    \item[2)] Via the PAP, an out-of-band, manual request is made to an intermediate, requesting to be listed as a federation member. 
    \item[3)] Via the PAP, the OP is bootstrapped. Configurations include domain name, keys, and immediate superiors. 
    \item[4)] The PAP configures the Workload with the TAs. 
    \item[5)] The workload requests a federation-scoped access token from the OP. Ambient provisioning is not sufficient, as it would give the workload an identity credential that does not have the right audience to be used to talk to a federation member. This contrasts with WIF, where an ambient-provisioned token can be sent directly to the other domain's STS.  
    \item[6)] Workload B initiates communication with Workload A. The locally issued, federation-scoped credential (obtained in step 5) is included in the request. As per OpenID Federation 1.0, how Workload A's endpoint is discovered is out of scope.
    \item[7)] Workload A fetches Domain B's Entity Statement, obtaining the signing keys to validate the credential received in step 6. Keys may be included directly in the entity configuration or fetched with a second request to the JWKS URI.    
    \item[8)] Step 7 alone is not sufficient to trust Workload B. Trust must be anchored to a TA. The intermediate authority that signed the entity statement (fetched in step 7) must be queried. It will return a list of valid subordinate entities in which B's OP must appear. Moreover, the entity statement of the intermediate is fetched, and it must be signed by a trusted TA.
    If successful, Workload A authorizes B's request received in step 6.
\end{enumerate}

\bibliographystyle{IEEEtran}  
\bibliography{references} 

@misc{gilman_harding_2021_spiffe,
author = {Gilman, Evan and Harding, Andrew},
title  = {{SPIFFE}: In Theory and in Practice},
howpublished = {CNCF [Cloud Native Computing Foundation] YouTube Channel},
month  = oct,
year   = {2021},
url    = {https://www.youtube.com/watch?v=DXE6CDJjDV4},
note   = {Presented at KubeCon + CloudNativeCon North America 2021. Accessed: May 18,
2026}
}

@misc{anthropic-wif,
author = {{Anthropic}},
title  = {Workload Identity Federation},
url    =
{https://platform.claude.com/docs/en/manage-claude/workload-identity-federation},
year   = {2026},
note   = {Accessed: 2026-05-21}
}

@misc{k8s-sa-token-projection,
author = {{The Kubernetes Authors}},
title  = {Configure Service Accounts for Pods: Service Account Token Volume
Projection},
url    = {https://kubernetes.io/docs/tasks/configure-pod-container/configure-service-account},
year   = {2026},
note   = {Accessed: 2026-05-21}
}

@misc{gcp-metadata,
author = {{Google Cloud}},
title  = {{VM} Metadata Server Overview},
url    = {https://cloud.google.com/compute/docs/metadata/overview},
year   = {2026},
note   = {Accessed: 2026-05-21}
}

@misc{aws-trust-policy,
author = {{Amazon Web Services}},
title  = {{IAM} Roles: Terms and Concepts --- Trust Policy},
url    =
{https://docs.aws.amazon.com/IAM/latest/UserGuide/id_roles_terms-and-concepts.html},
year   = {2026},
note   = {Accessed: 2026-05-21}
}

@misc{gcp-wif-pool-provider,
author = {{Google Cloud}},
title  = {Workload Identity Federation: Configure Providers},
url    = {https://docs.cloud.google.com/iam/docs/workload-identity-federation-with-oth
er-providers},
year   = {2026},
note   = {Accessed: 2026-05-21}
}

@misc{rfc6749,
    author = {Hardt, D.},
    title  = {{The OAuth 2.0 Authorization Framework}},
    howpublished = {RFC 6749},
    publisher = {IETF},
    year   = {2012},
    month  = oct,
    url    = {https://www.rfc-editor.org/rfc/rfc6749.html},
    doi    = {10.17487/RFC6749}
  }

@misc{oidc-core,
    author = {Sakimura, N. and Bradley, J. and Jones, M. and de Medeiros, B. and
  Mortimore, C.},
    title  = {{OpenID Connect Core 1.0 incorporating errata set 2}},
    howpublished = {OpenID Foundation},
    year   = {2023},
    url    = {https://openid.net/specs/openid-connect-core-1_0.html},
    note   = {Accessed: 2026-05-21}
  }

@techreport{nist-abac,
    author = {Hu, Vincent C. and Ferraiolo, David and Kuhn, Rick and Schnitzer, Adam and
  Sandlin, Kenneth and Miller, Robert and Scarfone, Karen},
    title  = {{Guide to Attribute Based Access Control ({ABAC}) Definition and
  Considerations}},
    number = {NIST Special Publication 800-162},
    institution = {National Institute of Standards and Technology},
    year   = {2014},
    url    = {https://doi.org/10.6028/NIST.SP.800-162},
    doi    = {10.6028/NIST.SP.800-162}
  }

@misc{oidf_about,
  author       = {{OpenID Foundation}},
  title        = {About {OpenID}},
  year         = {2023},
  howpublished = {\url{https://openid.net/foundation/}},
  note         = {Accessed: 2026-05-30}
}

@techreport{3gpp-s3-262608,
    author      = {{Deutsche Telekom} and {US NSA} and {Lenovo} and {Ericsson} and {Huawei} and {HiSilicon} and {China Mobile} and {Boost Mobile Network}},
    title       = {New Key Issue: Linking {NF} Authentication to
                   Authorization in {6G}},
    institution = {3rd Generation Partnership Project (3GPP)},
    type        = {Tdoc, pCR},
    number      = {S3-262608},
    year        = {2026},
    month       = may,
    address     = {Dalian, China},
    url         = {https://www.3gpp.org/ftp/tsg_sa/WG3_Security/TSGS3_128_Dalian/Docs/S3-262608.zip}
  }

@article{
    ayaz2021fhir,
  author  = {Ayaz, Muhammad and Pasha, Muhammad F. and Alzahrani, Mohammed Y. and Budiarto, Rahmat and Stiawan, Deris},
  title   = {The Fast Health Interoperability Resources ({FHIR}) Standard: Systematic Literature Review of Implementations, Applications, Challenges and Opportunities},
  journal = {JMIR Medical Informatics},
  year    = {2021},
  volume  = {9},
  number  = {7},
  pages   = {e21929},
  doi     = {10.2196/21929},
  url     = {https://doi.org/10.2196/21929}
}

@article{dinhle2019wearable,
  author  = {Dinh-Le, Catherine and Chuang, Rachel and Chokshi, Sara and Mann, Devin},
  title   = {Wearable Health Technology and Electronic Health Record Integration: Scoping Review and Future Directions},
  journal = {JMIR mHealth and uHealth},
  year    = {2019},
  volume  = {7},
  number  = {9},
  pages   = {e12861},
  doi     = {10.2196/12861},
  url     = {https://doi.org/10.2196/12861}
}

@techreport{fda2022precert,
  author      = {{U.S. Food and Drug Administration}},
  title       = {The Software Precertification ({Pre-Cert}) Pilot Program: Tailored Total Product Lifecycle Approaches and Key Findings},
  institution = {U.S. Food and Drug Administration, Digital Health Center of Excellence},
  year        = {2022},
  month       = sep,
  type        = {Final Report},
  url         = {https://www.fda.gov/medical-devices/digital-health-center-excellence/digital-health-software-precertification-pre-cert-pilot-program}
}

@article{crigger2019making,
  author  = {Crigger, Elliott and Khoury, Christopher},
  title   = {Making Policy on Augmented Intelligence in Health Care},
  journal = {AMA Journal of Ethics},
  volume  = {21},
  number  = {2},
  pages   = {E188--191},
  year    = {2019},
  doi     = {10.1001/amajethics.2019.188},
  url     = {https://journalofethics.ama-assn.org/article/making-policy-augmented-intelligence-health-care/2019-02}
}

@online{google2025a2a,
    author  = {{Google for Developers}},
    title   = {Announcing the {Agent2Agent} Protocol ({A2A}): A New Era of Agent Interoperability},
    date    = {2025-04-09},
    url     = {https://developers.googleblog.com/en/a2a-a-new-era-of-agent-interoperability/},
    urldate = {2026-06-04}
  }

@misc{ehtesham2025interopsurvey,
    author       = {Ehtesham, Abul and Singh, Aditi and Gupta, Gaurav Kumar and Kumar, Saket},
    title        = {A Survey of Agent Interoperability Protocols: {Model} {Context} {Protocol} ({MCP}), {Agent} {Communication} {Protocol} ({ACP}), {Agent-to-Agent} {Protocol}
  ({A2A}), and {Agent} {Network} {Protocol} ({ANP})},
    year         = {2025},
    eprint       = {2505.02279},
    archivePrefix= {arXiv},
    primaryClass = {cs.AI},
    url          = {https://arxiv.org/abs/2505.02279}
  }

@inproceedings{garzon2024beyondcertificates,
    author    = {Rodriguez Garzon, Sandro and Dinh Tuan, Hai and Mora Martinez, Maria and K{\"u}pper, Axel and Einsiedler, Hans Joachim and Schneider, Daniela},
    title     = {Beyond Certificates: {6G}-ready Access Control for the Service-Based Architecture with Decentralized Identifiers and Verifiable Credentials},
    booktitle = {2024 Joint European Conference on Networks and Communications \& {6G} Summit (EuCNC/6G Summit)},
    pages     = {830--835},
    year      = {2024},
    doi       = {10.1109/EuCNC/6GSummit60053.2024.10597085},
    url       = {https://arxiv.org/abs/2310.19366}
  }

@misc{spiffe_scaling_spire,
    author       = {{SPIFFE Project}},
    title        = {Scaling SPIRE},
    howpublished = {\url{https://spiffe.io/docs/latest/planning/scaling_spire/}},
    note         = {Accessed: 2026-06-05}
}

@misc{spidcie_oidf,
    author       = {{Developers Italia}},
    title        = {{spid-cie-oidc-django}: The {SPID/CIE} {OpenID} Connect Federation {SDK}},
    howpublished = {GitHub repository},
    year         = {2024},
    url          = {https://github.com/italia/spid-cie-oidc-django},
    urldate      = {2026-06-05},
    note         = {Italy's SPID/CIE national digital identity on OpenID Federation; led by G. De Marco}
  }

@misc{raidiam_oidf,
    author       = {{Raidiam}},
    title        = {{OpenID} Federation --- The Missing Link for Scalable Trust in Data, {AI}, and Wallet Ecosystems},
    howpublished = {Raidiam Developers Blog},
    year         = {2025},
    url          = {https://www.raidiam.com/developers/blog/open-id-federation-the-missing-link-for-scalable-trust-in-data-ai-and-wallet-ecosystems},
    urldate      = {2026-06-05},
    note         = {Raidiam operates the Open Finance Brasil trust directory}
  }

@online{openid_federation,
    title   = {{OpenID} {Federation} 1.0},
    author  = {Hedberg, Roland and Jones, Michael B. and Solberg, Andreas {\AA}kre and Bradley, John and {De Marco}, Giuseppe and Dzhuvinov, Vladimir},
    organization = {OpenID Foundation},
    date    = {2026-02-17},
    version = {Final},
    url     = {https://openid.net/specs/openid-federation-1_0-final.html},
    urldate = {2026-06-05}
  }

@techreport{rfc8693,
    author      = {Jones, M. and Nadalin, A. and Campbell, B. and Bradley, J. and Mortimore, C.},
    title       = {{OAuth 2.0 Token Exchange}},
    type        = {RFC},
    number      = {8693},
    institution = {Internet Engineering Task Force (IETF)},
    year        = {2020},
    month       = jan,
    doi         = {10.17487/RFC8693},
    url         = {https://www.rfc-editor.org/rfc/rfc8693},
    note        = {Proposed Standard}
  }

@online{gcp_wif,
    author       = {{Google Cloud}},
    title        = {Workload Identity Federation},
    organization = {Google Cloud IAM Documentation},
    url          = {https://cloud.google.com/iam/docs/workload-identity-federation},
    urldate      = {2026-06-05}
  }

@book{
    cellularBook,
    title={Cellular: an economic and business history of the international mobile-phone industry},
    author={Garcia-Swartz, Daniel D and Campbell-Kelly, Martin},
    year={2022},
    publisher={MIT Press}
}

@misc{
    googleTelefnicaCase,
    author = {},
    title = {Telefónica Case Study},
    howpublished = {\url{https://cloud.google.com/customers/telefonica}},
    year = {},
    note = {[Accessed 16-12-2025]},
}

@misc{
    awsWavelength,
    author = {},
    title = {{AWS Wavelength}},
    howpublished = {\url{https://aws.amazon.com/wavelength/}},
    year = {},
    note = {[Accessed 16-12-2025]},
}

@misc{rhapsodyReasonsInteroperability,
	author = {},
	title = {4 {R}easons {W}hy {E}{H}{R} {I}nteroperability is a {M}ess (and {H}ow to {F}ix {I}t) --- rhapsody.health},
	howpublished = {\url{https://rhapsody.health/blog/reasons-ehr-interoperability-is-a-mess-and-how-to-fix-it/}},
	year = {},
	note = {[Accessed 30-03-2026]},
}

@misc{kffhealthnewsDeath1000,
	author = {Fred Schulte, Erika Fry, Fortune},
	title = {{D}eath {B}y 1,000 {C}licks: {W}here {E}lectronic {H}ealth {R}ecords {W}ent {W}rong - {K}{F}{F} {H}ealth {N}ews --- kffhealthnews.org},
	howpublished = {\url{https://kffhealthnews.org/news/death-by-a-thousand-clicks/}},
	year = {},
	note = {[Accessed 30-03-2026]},
}

@misc{participatorymedicine,
	author = {Ferguson et al.},
	title = {e-Patients: How They Can Help Us Heal Health Care},
	howpublished = {\url{https://participatorymedicine.org/e-Patients_White_Paper.pdf}},
	year = {},
	note = {[Accessed 30-03-2026]},
}

@misc{richards2013let,
  title={Let the patient revolution begin},
  author={Richards, Tessa and Montori, Victor M and Godlee, Fiona and Lapsley, Peter and Paul, Dave},
  journal={Bmj},
  volume={346},
  year={2013},
  publisher={British Medical Journal Publishing Group}
}

@article{thomas2025artificial,
  title={Artificial intelligence in modern clinical practice},
  author={Thomas, Krupa Sara and Edpuganti, Sudeep and Puthooran, Divina Mariya and Thomas, Angela and Joy, Angel and Latheef, Shifna},
  journal={Medicine International},
  volume={6},
  number={1},
  pages={5},
  year={2025},
  publisher={DA Spandidos}
}

@article{johnson2021precision,
  title={Precision medicine, AI, and the future of personalized health care},
  author={Johnson, Kevin B and Wei, Wei-Qi and Weeraratne, Dilhan and Frisse, Mark E and Misulis, Karl and Rhee, Kyu and Zhao, Juan and Snowdon, Jane L},
  journal={Clinical and translational science},
  volume={14},
  number={1},
  pages={86--93},
  year={2021},
  publisher={Wiley Online Library}
}

@article{
admin_e_health,
author = {Lavoie-Gagne, Ophelie and Woo, Joshua J. and Williams III, Riley J. and Nwachukwu, Benedict U. and Kunze, Kyle N. and Ramkumar, Prem N.},
title = {Artificial Intelligence as a Tool to Mitigate Administrative Burden, Optimize Billing, Reduce Insurance- and Credentialing-Related Expenses, and Improve Quality Assurance Within Health Care Systems},
journal = {Arthroscopy},
volume = {41},
number = {8},
pages = {3270-3275},
doi = {https://doi.org/10.1016/j.arthro.2025.02.038},
url = {https://arthroscopyjournals.onlinelibrary.wiley.com/doi/abs/10.1016/j.arthro.2025.02.038},
eprint = {https://arthroscopyjournals.onlinelibrary.wiley.com/doi/pdf/10.1016/j.arthro.2025.02.038},
year = {2025}
}

@misc{healthdatamanagementPatientMatching,
	author = {},
	title = {{W}hy patient matching is a key challenge in achieving interoperability - {H}ealth {D}ata {M}anagement --- healthdatamanagement.com},
	howpublished = {\url{https://www.healthdatamanagement.com/articles/why-patient-matching-is-a-key-challenge-in-achieving-interoperability?id=135986}},
	year = {},
	note = {[Accessed 31-03-2026]},
}

@misc{garzon2025aiagentsdecentralizedidentifiers,
      title={AI Agents with Decentralized Identifiers and Verifiable Credentials}, 
      author={Sandro Rodriguez Garzon and Awid Vaziry and Enis Mert Kuzu and Dennis Enrique Gehrmann and Buse Varkan and Alexander Gaballa and Axel Küpper},
      year={2025},
      eprint={2511.02841},
      archivePrefix={arXiv},
      primaryClass={cs.CR},
      url={https://arxiv.org/abs/2511.02841}, 
}

@misc{anp_whitepaper,
    title         = {{Agent Network Protocol} Technical White Paper},
    author        = {Chang, Gaowei and Lin, Eidan and Yuan, Chengxuan and Cai, Rizhao and Chen, Binbin and Xie, Xuan and Zhang, Yin},
    year          = {2025},
    eprint        = {2508.00007},
    archivePrefix = {arXiv},
    primaryClass  = {cs.AI},
    url           = {https://arxiv.org/abs/2508.00007}
  }

@misc{2025programmable,
    title         = {Programmable Governance for Group-Controlled Decentralized
  Identifiers},
    author        = {Segat, Carlo and Rodriguez Garzon, Sandro and K\"upper, Axel},
    year          = {2025},
    eprint        = {2507.06001},
    archivePrefix = {arXiv},
    primaryClass  = {cs.CR},
    url           = {https://arxiv.org/abs/2507.06001}
  }

@article{longpre2023dataprovenance,
    title   = {The Data Provenance Initiative: A Large Scale Audit of Dataset Licensing \&
  Attribution in AI},
    author  = {Longpre, Shayne and Mahari, Robert and Chen, Anthony and Obeng-Marnu, Naana
  and Sileo, Damien and Brannon, William and Muennighoff, Niklas and Khazam, Nathan and
  Kabbara, Jad and Perisetla, Kartik and Wu, Xinyi and Shippole, Enrico and Bollacker, Kurt
  and Wu, Tongshuang and Villa, Luis and Pentland, Sandy and Hooker, Sara},
    journal = {arXiv preprint arXiv:2310.16787},
    year    = {2023}
  }

@inproceedings{sevilla2022compute,
    title        = {Compute Trends Across Three Eras of Machine Learning},
    author       = {Sevilla, Jaime and Heim, Lennart and Ho, Anson and Besiroglu, Tamay and
  Hobbhahn, Marius and Villalobos, Pablo},
    booktitle    = {2022 International Joint Conference on Neural Networks (IJCNN)},
    pages        = {1--8},
    year         = {2022},
    organization = {IEEE}
  }

@misc{nvidia2025q4,
    author       = {{NVIDIA Corporation}},
    title        = {NVIDIA Announces Financial Results for Fourth Quarter and Fiscal Year
  2025},
    howpublished = {Press release, Form 8-K, U.S. Securities and Exchange Commission},
    year         = {2025},
    url          =
  {https://www.sec.gov/Archives/edgar/data/0001045810/000104581025000021/q4fy25pr.htm}
  }

@article{arrietaibarra2018data,
    title   = {Should We Treat Data as Labor? Moving beyond ``Free''},
    author  = {Arrieta-Ibarra, Imanol and Goff, Leonard and Jim\'enez-Hern\'andez, Diego
  and Lanier, Jaron and Weyl, E. Glen},
    journal = {AEA Papers and Proceedings},
    volume  = {108},
    pages   = {38--42},
    year    = {2018}
  }

@article{ganesan2023sidelink,
    author  = {Ganesan, Karthikeyan and Loehr, Joachim and Choi, Hyung-Nam and others},
    title   = {{5G} Advanced: Sidelink Evolution},
    journal = {IEEE Communications Standards Magazine},
    volume  = {7},
    number  = {1},
    pages   = {58--63},
    year    = {2023},
    doi     = {10.1109/MCOMSTD.0007.2200057},
    url     = {https://doi.org/10.1109/MCOMSTD.0007.2200057}
  }

@misc{csaA2AThreatModel,
    author       = {{Cloud Security Alliance}},
    title        = {Threat Modeling Google's {A2A} Protocol with the {MAESTRO} Framework},
    howpublished = {Cloud Security Alliance Blog},
    year         = {2025},
    month        = apr,
    url          = {https://cloudsecurityalliance.org/blog/2025/04/30/threat-modeling-goog
  le-s-a2a-protocol-with-the-maestro-framework},
    note         = {Accessed: 2026-06-10}
  }

@article{babakian2022retrospective,
    author  = {Babakian, Andrew and Monclus, Pere and Braun, Robin and Lipman, Justin},
    title   = {A Retrospective on Workload Identifiers: From Data Center to Cloud-Native
  Networks},
    journal = {IEEE Access},
    year    = {2022},
    volume  = {10},
    pages   = {105518--105527},
    doi     = {10.1109/ACCESS.2022.3211293},
    url     = {https://doi.org/10.1109/ACCESS.2022.3211293}
  }

@techreport{rfc9334,
    author      = {Birkholz, H. and Thaler, D. and Richardson, M. and Smith, N. and Pan, W.},
    title       = {{Remote ATtestation procedureS (RATS) Architecture}},
    type        = {RFC},
    number      = {9334},
    institution = {Internet Engineering Task Force (IETF)},
    year        = {2023},
    month       = jan,
    doi         = {10.17487/RFC9334},
    url         = {https://www.rfc-editor.org/rfc/rfc9334}
  }

@online{spiffe_concepts,
    author       = {{SPIFFE}},
    title        = {{SPIRE} Concepts},
    organization = {Cloud Native Computing Foundation},
    url          = {https://spiffe.io/docs/latest/spire-about/spire-concepts/},
    urldate      = {2026-08-23}
  }

@online{keylime,
    author       = {{Keylime}},
    title        = {{Keylime}: Remote Boot Attestation and Runtime Integrity Measurement},
    organization = {Cloud Native Computing Foundation},
    url          = {https://keylime.dev/},
    urldate      = {2026-08-23}
  }

@online{veraison,
    author       = {{Project Veraison}},
    title        = {{Veraison}: Attestation Verification Services},
    organization = {Confidential Computing Consortium, Linux Foundation},
    url          = {https://www.veraison-project.org/},
    urldate      = {2026-08-23}
  }

@online{keylime_spire,
    author       = {Bulwahn, Lukas and Robb, Anderson},
    title        = {{SPIFFE/SPIRE} and {Keylime}: Software Identity based on Secure Machine State},
    organization = {Red Hat Emerging Technologies},
    year         = {2025},
    month        = jan,
    url          = {https://next.redhat.com/2025/01/24/spiffe-spire-and-keylime-software-identity-based-on-secure-machine-state/},
    urldate      = {2026-08-23}
  }

@techreport{wimse_wit_attestation,
    author      = {Liu, D. and Zhu, H. and Jin, J.},
    title       = {{Carrying Remote Attestation Evidence in Workload Identity Tokens (WIT)}},
    type        = {Internet-Draft},
    number      = {draft-liu-wimse-wit-attestation-00},
    institution = {Internet Engineering Task Force (IETF)},
    year        = {2026},
    month       = mar,
    url         = {https://datatracker.ietf.org/doc/draft-liu-wimse-wit-attestation/}
  }

@techreport{wimse_transitive,
    author      = {Krishnan, R. and Prasad, A. and Lopez, D. and Addepalli, S.},
    title       = {{Transitive Attestation for Sovereign Workloads: A WIMSE Profile}},
    type        = {Internet-Draft},
    number      = {draft-mw-wimse-transitive-attestation-00},
    institution = {Internet Engineering Task Force (IETF)},
    year        = {2026},
    month       = feb,
    url         = {https://datatracker.ietf.org/doc/draft-mw-wimse-transitive-attestation/}
  }

@techreport{rats_federated_infra,
    author      = {Lu, D.},
    title       = {{Federated Infrastructure Layer for Cross-Platform Device and Application Attestation}},
    type        = {Internet-Draft},
    number      = {draft-lu-rats-federated-attestation-infrastructure-00},
    institution = {Internet Engineering Task Force (IETF)},
    year        = {2026},
    month       = may,
    url         = {https://datatracker.ietf.org/doc/draft-lu-rats-federated-attestation-infrastructure/}
  }

\end{document}